\documentclass[sigconf,table]{acmart}

\AtBeginDocument{%
  \providecommand\BibTeX{{%
    \normalfont B\kern-0.5em{\scshape i\kern-0.25em b}\kern-0.8em\TeX}}}

\copyrightyear{2022}
\acmYear{2022}
\acmDOI{XXXXXXX.XXXXXXX}
\acmConference[ICPC 2022]{30th IEEE/ACM International Conference on Program Comprehension}{May 16--17,
  2022}{Pittsburgh, PA, USA}

\usepackage{tcolorbox}
\usepackage{xcolor}
\usepackage{booktabs}
\usepackage{enumitem}
\usepackage{caption}
\usepackage{subcaption}
\usepackage{lineno,hyperref}
\usepackage[linesnumbered,boxruled]{algorithm2e}
\usepackage{amsmath,colortbl}
\usepackage{flushend}
\usepackage{graphicx}
\usepackage{dblfloatfix}
\usepackage{textcomp}
\usepackage{url}
\usepackage{multirow}
\usepackage[colorinlistoftodos]{todonotes}
\usepackage{listings}
\usepackage{comment}
\setlist[description]{leftmargin=\parindent,labelindent=\parindent}

\newlist{UP}{enumerate}{1}
\setlist[UP]{label=U\textsubscript{\arabic*}:}
\newlist{RP}{enumerate}{1}
\setlist[RP]{label=R\textsubscript{\arabic*}:}
\newlist{CP}{enumerate}{1}
\setlist[CP]{label=C\textsubscript{\arabic*}:}
\newlist{FP}{enumerate}{1}
\setlist[FP]{label=F\textsubscript{\arabic*}:}

\usepackage[percent]{overpic}%
\usepackage{tikz}
\usepackage{moresize}
\colorlet{punct}{red!60!black}
\definecolor{background}{HTML}{EEEEEE}
\definecolor{delim}{RGB}{20,105,176}
\colorlet{numb}{magenta!60!black}

\modulolinenumbers[5]

\lstdefinelanguage{json}{
    basicstyle=\scriptsize,
    numberstyle=\scriptsize,
    stepnumber=1,
    numbersep=5pt,
    showstringspaces=false,
    breaklines=true,
    frame=lines,
    backgroundcolor=\color{background},
    literate=
     *{0}{{{\color{numb}0}}}{1}
      {1}{{{\color{numb}1}}}{1}
      {2}{{{\color{numb}2}}}{1}
      {3}{{{\color{numb}3}}}{1}
      {4}{{{\color{numb}4}}}{1}
      {5}{{{\color{numb}5}}}{1}
      {6}{{{\color{numb}6}}}{1}
      {7}{{{\color{numb}7}}}{1}
      {8}{{{\color{numb}8}}}{1}
      {9}{{{\color{numb}9}}}{1}
      {:}{{{\color{punct}{:}}}}{1}
      {,}{{{\color{punct}{,}}}}{1}
      {\{}{{{\color{delim}{\{}}}}{1}
      {\}}{{{\color{delim}{\}}}}}{1}
      {[}{{{\color{delim}{[}}}}{1}
      {]}{{{\color{delim}{]}}}}{1},
}

\definecolor{maroon}{cmyk}{0, 0.87, 0.68, 0.32}
\definecolor{halfgray}{gray}{0.55}
\definecolor{ipython_frame}{RGB}{207, 207, 207}
\definecolor{ipython_bg}{RGB}{247, 247, 247}
\definecolor{ipython_red}{RGB}{186, 33, 33}
\definecolor{ipython_green}{RGB}{0, 128, 0}
\definecolor{ipython_cyan}{RGB}{64, 128, 128}
\definecolor{ipython_purple}{RGB}{170, 34, 255}

\lstdefinelanguage{python}{
    morekeywords={access,and,break,class,continue,def,del,elif,else,except,exec,finally,for,from,global,if,import,in,is,lambda,not,or,pass,print,raise,return,try,while},
    morekeywords=[2]{abs,all,any,basestring,bin,bool,bytearray,callable,chr,classmethod,cmp,compile,complex,delattr,dict,dir,divmod,enumerate,eval,execfile,file,filter,float,format,frozenset,getattr,globals,hasattr,hash,help,hex,id,input,int,isinstance,issubclass,iter,len,list,locals,long,map,max,memoryview,min,next,object,oct,open,ord,pow,property,range,raw_input,reduce,reload,repr,reversed,round,set,setattr,slice,sorted,staticmethod,str,sum,super,tuple,type,unichr,unicode,vars,xrange,zip,apply,buffer,coerce,intern},
    sensitive=true,
    morecomment=[l]\#,
    morestring=[b]',
    morestring=[b]",
    morestring=[s]{'''}{'''},
    morestring=[s]{"""}{"""},
    morestring=[s]{r'}{'},
    morestring=[s]{r"}{"},
    morestring=[s]{r'''}{'''},
    morestring=[s]{r"""}{"""},
    morestring=[s]{u'}{'},
    morestring=[s]{u"}{"},
    morestring=[s]{u'''}{'''},
    morestring=[s]{u"""}{"""},
    literate=
    {á}{{\'a}}1 {é}{{\'e}}1 {í}{{\'i}}1 {ó}{{\'o}}1 {ú}{{\'u}}1
    {Á}{{\'A}}1 {É}{{\'E}}1 {Í}{{\'I}}1 {Ó}{{\'O}}1 {Ú}{{\'U}}1
    {à}{{\`a}}1 {è}{{\`e}}1 {ì}{{\`i}}1 {ò}{{\`o}}1 {ù}{{\`u}}1
    {À}{{\`A}}1 {È}{{\'E}}1 {Ì}{{\`I}}1 {Ò}{{\`O}}1 {Ù}{{\`U}}1
    {ä}{{\"a}}1 {ë}{{\"e}}1 {ï}{{\"i}}1 {ö}{{\"o}}1 {ü}{{\"u}}1
    {Ä}{{\"A}}1 {Ë}{{\"E}}1 {Ï}{{\"I}}1 {Ö}{{\"O}}1 {Ü}{{\"U}}1
    {â}{{\^a}}1 {ê}{{\^e}}1 {î}{{\^i}}1 {ô}{{\^o}}1 {û}{{\^u}}1
    {Â}{{\^A}}1 {Ê}{{\^E}}1 {Î}{{\^I}}1 {Ô}{{\^O}}1 {Û}{{\^U}}1
    {œ}{{\oe}}1 {Œ}{{\OE}}1 {æ}{{\ae}}1 {Æ}{{\AE}}1 {ß}{{\ss}}1
    {ç}{{\c c}}1 {Ç}{{\c C}}1 {ø}{{\o}}1 {å}{{\r a}}1 {Å}{{\r A}}1
    {€}{{\EUR}}1 {£}{{\pounds}}1
    {^}{{{\color{ipython_purple}\^{}}}}1
    {=}{{{\color{ipython_purple}=}}}1
    {+}{{{\color{ipython_purple}+}}}1
    {*}{{{\color{ipython_purple}$^\ast$}}}1
    {/}{{{\color{ipython_purple}/}}}1
    {+=}{{{+=}}}1
    {-=}{{{-=}}}1
    {*=}{{{$^\ast$=}}}1
    {/=}{{{/=}}}1,
    literate=
    *{-}{{{\color{ipython_purple}-}}}1
     {?}{{{\color{ipython_purple}?}}}1,
    identifierstyle=\color{black}\ttfamily,
    commentstyle=\color{ipython_cyan}\ttfamily,
    stringstyle=\color{ipython_red}\ttfamily,
    keepspaces=true,
    showspaces=false,
    showstringspaces=false,
    rulecolor=\color{ipython_frame},
    numberstyle=\tiny\color{halfgray},
    backgroundcolor=\color{ipython_bg},
    basicstyle=\scriptsize,
    keywordstyle=\color{ipython_green}\ttfamily,
}

\def\BibTeX{{\rm B\kern-.05em{\sc i\kern-.025em b}\kern-.08em
    T\kern-.1667em\lower.7ex\hbox{E}\kern-.125emX}}

\begin{document}
\begin{sloppy}

\newcommand\redtodo[1]{\textcolor{red}{#1}}
\newcommand\bodin[1]{{\textcolor{blue}{#1}}}
\newcommand\asia[1]{{\textcolor{blue}{(#1)}}}
\newcommand\raula[1]{{\textcolor{blue}{(#1)}}}
\newcommand\ishio[1]{{\textcolor{blue}{(#1)}}}
\newcommand\assign[1]{{\colorbox{yellow}{\textcolor{red}{@#1}}}}

\newcommand{\RqOne}{RQ1}
\newcommand{\RqOneSen}{(\RqOne) Do Pythonic idioms improve the memory usage?}
\newcommand{\RqTwo}{RQ2}
\newcommand{\RqTwoSen}{(\RqTwo) Do Pythonic idioms improve the execution time?}

\newcommand{\inputSize}{Run}

\title{Does Coding in Pythonic Zen Peak Performance? \\ Preliminary Experiments of Nine Pythonic Idioms at Scale}


\author{Pattara Leelaprute}
\affiliation{%
  \institution{Faculty of Engineering, Kasetsart University}
  \country{}
}
\email{pattara.l@ku.ac.th}

\author{Bodin Chinthanet}
\affiliation{%
  \institution{Nara Institute of Science and Technology}
  \country{}
}
\email{bodin.ch@is.naist.jp}

\author{Supatsara Wattanakriengkrai}
\affiliation{%
  \institution{Nara Institute of Science and Technology}
  \country{}
}
\email{wattanakri.supatsara.ws3@is.naist.jp}

\author{Raula Gaikovina Kula}
\affiliation{%
  \institution{Nara Institute of Science and Technology}
  \country{}
}
\email{raula-k@is.naist.jp}

\author{Pongchai Jaisri}
\affiliation{%
  \institution{Faculty of Engineering, Kasetsart University}
  \country{}
}
\email{pongchai.j@ku.th}

\author{Takashi Ishio}
\affiliation{%
  \institution{Nara Institute of Science and Technology}
  \country{}
}
\email{ishio@is.naist.jp}

\begin{abstract}

In the field of data science, and for academics in general, the Python programming language is a popular choice,  mainly because of its libraries for storing, manipulating, and gaining insight from data. Evidence includes the versatile set of machine learning, data visualization, and manipulation packages used for the ever-growing size of available data. The \textit{ Zen of Python} is a set of guiding design principles that developers use to write acceptable and elegant Python code. Most principles revolve around simplicity. However, as the need to compute large amounts of data, performance has become a necessity for the Python programmer.
The new idea in this paper is to confirm whether writing the Pythonic way peaks performance at scale. As a starting point, we conduct a set of preliminary experiments to evaluate nine Pythonic code examples by comparing the performance of both Pythonic and Non-Pythonic code snippets. Our results reveal that writing in Pythonic idioms may save memory and time.
We show that incorporating list comprehension, generator expression, zip, and itertools.zip\_longest idioms can save up to 7,000 MB and up to 32.25 seconds. The results open more questions on how they could be utilized in a real-world setting. The replication package includes all scripts, and the results are available at \url{https://doi.org/10.5281/zenodo.5712349}

\end{abstract}


\maketitle

\section{Introduction}
\label{sec:introduction}
The Python programming language is one of the most popular and versatile programming language to learn, especially for professionals and students that deal with data analysis and in the field of data analytics.
According to an anaconda report in 2021 \footnote{\url{https://www.anaconda.com/state-of-data-science-2021}},  63\% of respondents said they used Python frequently or always while 71\% of educators said they're teaching machine learning and data science with Python, which has become popular because of its ease of use and easy learning curve, while 88\% of students said they were being taught Python in preparation to enter the data science/machine learning field. 
Furthermore, in 2021, Python was ranked at number one on the TIOBE Index based on the number of skilled engineers worldwide, courses, and third-party vendors.\footnote{\url{https://www.tiobe.com/tiobe-index/}}
Furthermore, it is ranked as being the second most popular on GitHub.\footnote{\url{https://octoverse.github.com/\#top-languages-over-the-years}}

Like other programming communities, the Python community has over time developed their own agreed consensus on how to write Python \citep{farooq:2021}.
From programmer blogs, mailing lists, and down to official documentation, the term ‘Pythonic’ is used liberally by the Python community.
Hence, the \textit{Zen of Python} is a set of 19 design principles \citep{PEP20:online} for writing computer programs that influence the design of the Python programming language \citep{PythonGlossary:online}.
The principles include concepts such as ``Beautiful is better than ugly", ``Explicit is better than implicit", and ``Simple is better than complex".
Intuitively, Non-Pythonic code may look imperfect to an experienced Python developer.
\citet{farooq:2021}, shows that the scope of the term ‘Pythonic’ appears to go far beyond concrete source code, referring to a way of thinking about problems and potential solutions. 
Researchers have only begun to study how developers use Pythonic idioms.
Recent work is now concerned with the different aspects of Pythonic detection and usage \cite{Alexandru:Onward2018, Tatt:IWESEP2019, Phan-udom:ICSME2020, Merchante:2017, Orlov:2020}.

In recent times, the Python language is been utilized by both the researcher and practitioner for data analysis.
Python is not only popular in many different settings not only with software engineers as users, but with data scientists, and students alike for machine learning, visualization, and data analysis.
Evidence of this popularity can be the rise of Jupyter notebooks \cite{KGTorrentMSR21} and data repositories such as Kaggle \footnote{\url{https://www.kaggle.com/}}, 
and the rise of libraries (such as Scikit-learn,\footnote{\url{https://scikit-learn.org}} TensorFlow,\footnote{\url{https://www.tensorflow.org}} and Keras\footnote{\url{https://keras.io}}).
With the increase in software artifacts becoming available on sites such as GitHub, the Pythonic programmer need to handle size, complexity and memory constraints.

The new idea in this paper is to start preliminary investigations into the performance of Pythonic code in terms of memory and execution.
In a case study of nine Pythonic idioms (i.e., list comprehension, dict comprehension, generator expression, yield, lambda, defaultdict, deque, zip, zip\_longest),  we run a performance test under different input level settings (up to 100,000,000 integer instances) to test memory and execution usage.
Results indicate that some performance benefits are realized at larger computations for some idioms.
Encouraged by the results, our future plan includes empirical evaluation on real-world applications. 

\begin{table*}[b]
\centering
\caption{Summary statistics of median improvement of Pythonic code over Non-Pythonic code in terms of memory usages and execution times (MB | seconds) from Run = 1 to 9 (i.e., 10$^0$ to 10$^8$ integer instances). ``-'' refers to no improvement.}
\label{tab:mem_time_diff}
\scalebox{0.84}{
\begin{tabular}{@{}cccccccccc@{}}
\toprule
\multirow{2}{*}{} & \multicolumn{9}{c}{\textbf{Runs (MB|secs)}} \\ \cmidrule(l){2-10} 
 & \textbf{\textbf{$\Delta$} @1} & \textbf{$\Delta$ @2} & \textbf{$\Delta$ @3} & \textbf{$\Delta$ @4} & \textbf{$\Delta$ @5} & \textbf{$\Delta$ @6} & \textbf{$\Delta$ @7} & \textbf{$\Delta$ @8} & \textbf{$\Delta$ @9} \\ \midrule
List comprehension & 0.05 | - & - | - & - | - & 0.02 | - & - | - & 0.07 | 0.01 & 0.02 | 0.06 & 0.05 | 0.58 & 0.05 | 5.78 \\
Dict comprehension & 0.01 | - & - | - & - | - & - | - & 0.02 | - & 0.03 | - & - | 0.04 & - | 0.45 & - | 4.46 \\
Generator expression & - | - & 0.03 | - & - | - & 0.01 | - & 0.21 | - & 3.82 | - & 38.61 | 0.01 & 386.36 | 0.11 & 3,863.95 | 1.10 \\
yield & - | - & 0.03 | - & 0.04 | - & - | - & 0.19 | - & 4.10 | - & 38.92 | 0.02 & 386.58 | 0.17 & 3,864.13 | 1.70 \\
lambda & 0.02 | - & - | - & 0.05 | - & - | - & - | - & - | - & 0.02 | 0.01 & - | 0.05 & - | 0.53 \\
collections.defaultdict & - | - & 0.03 | - & - | - & - | - & 0.01 | - & - | - & - | - & - | - & - | - \\
collections.deque & 0.01 | - & - | - & - | - & - | - & 0.02 | - & 0.05 | - & - | 0.02 & - | 0.22 & - | 2.07 \\
zip & 0.04 | - & - | - & - | - & - | - & 0.28 | - & 3.61 | 0.01 & 34.90 | 0.12 & 348.50 | 1.13 & 3,481.89 | 11.02 \\
itertools.zip\_longest & 0.01 | - & - | - & 0.02 | - & - | - & 0.79 | - & 7.26 | 0.04 & 70.01 | 0.35 & 696.79 | 3.28 & 6,964.85 | 32.25 \\ \bottomrule
\end{tabular}
}
\end{table*}

\section{A Catalog of Nine Pythonic Idioms}

We target nine Pythonic idioms identified in prior works \cite{farooq:2021, Alexandru:Onward2018} and acknowledged to have performance benefits through a literature review and developer interviews.
We now explain each idiom and show its example.\footnote{These examples were synthesized from an online catalog of Pythonic idioms provided by \citet{Alexandru:Onward2018}:  \url{https://pythonic-examples.github.io}}

\textit{1. List comprehension} is a concise way of constructing new lists from other sequences, such as applying an operation to each element (similar to `map' in other languages).

\begin{lstlisting}[language=Python,
caption={},
label=code:userhome]
# Note: n is integer
result_list = [el for el in range(n)]
\end{lstlisting}

\textit{2. Dict comprehension} is similar to list comprehension but is used for constructing dictionaries, e.g., `Maps' in other languages.

\begin{lstlisting}[language=Python,
caption={},
label=code:userhome]
# Note: n is integer
dict_compr = {k: k*2 for k in range(n)}
\end{lstlisting}

\textit{3. Generator expression} is also similar to a list comprehension, but for lazily generating sequences.

\begin{lstlisting}[language=Python,
caption={},
label=code:userhome]
# Note: n is integer
result = sum((x*2 for x in range(n)))
\end{lstlisting}

\textit{4. yield.} creates generators instead of utilizing return. It is an easier way to write generators than to build an iterator (using the \_\_next\_\_ magic method).

\begin{lstlisting}[language=Python,
caption={},
label=code:userhome]
# Note: n is integer
def two_times_range(n):
    for it in range(n):
        yield it*2
\end{lstlisting}

\textit{5. lambda.} implements anonymous functions.

\begin{lstlisting}[language=Python,
caption={},
label=code:userhome]
# Note: n is integer
#       n_list is list of n integer elements
result = sum(map(lambda it: it*2 + 2, n_list))
\end{lstlisting}

\textit{6. collections.defaultdict.} is similar to a simple dictionary, but when accessing a non-existent key, a default value is added and returned in place of raising an exception.
\begin{lstlisting}[language=Python,
caption={},
label=code:userhome]
# Note: n is integer
#       n_list is list of n integer elements
from collections import defaultdict
def add_new_value_in_dict_pythonic(n_list):
    num_dict = defaultdict(int)
    for key in n_list:
        num_dict[key] -= key
    return num_dict
\end{lstlisting}

\textit{7. collections.deque} is a generalization of constructing stacks and queues.

\begin{lstlisting}[language=Python,
caption={},
label=code:userhome]
# Note: n is integer
from collections import deque
def add_new_value_in_queue_pythonic(n):
    num_queue = deque([])
    for num in range(n):
        if num % 4 == 0:
            num_queue.append(num)
        elif num % 4 == 1:
            num_queue.appendleft(num)
        elif num % 4 == 2:
            num_queue.pop()
        else:
            num_queue.popleft()
    return num_queue
\end{lstlisting}

\textit{8. zip. } accepts two or more iterables and returns a new iterator where each i-th element yielded is a tuple containing the i-th elements of each input.

\begin{lstlisting}[language=Python,
caption={},
label=code:userhome]
# Note: n is integer
#       a_list is list of n integer elements
#       b_list is list of n/2 integer instances
for a, b in zip(a_list, b_list):
    result = min(a, b)
\end{lstlisting}

\begin{table*}[h]
\centering
\caption{Summary statistics and comparison of memory usages (in MB) between Pythonic and Non-pythonic codes. This result is taken from \inputSize~= 9 (i.e., 10$^8$ integer instances). $\Delta$ is improvement of Pythonic code over Non-Pythonic code (based on median).} 
\label{tab:result_mem_stat}
\scalebox{0.82}{
\begin{tabular}{@{}crrrrrrccc@{}}
\toprule
 & \multicolumn{3}{c}{\textbf{Pythonic}} & \multicolumn{3}{c}{\textbf{Non-Pythonic}} & \multicolumn{1}{c}{\multirow{2}{*}{\textbf{$\Delta$}}} & \multirow{2}{*}{\textbf{t-test}} & \multirow{2}{*}{\textbf{Cohen's d}} \\ \cmidrule(r){2-7}
 & \multicolumn{1}{c}{\textbf{Median}} & \multicolumn{1}{c}{\textbf{Mean}} & \multicolumn{1}{c}{\textbf{SD}} & \multicolumn{1}{c}{\textbf{Median}} & \multicolumn{1}{c}{\textbf{Mean}} & \multicolumn{1}{c}{\textbf{SD}} & & (p-value $<$ 0.001)  &  \\ \midrule
List comprehension & \cellcolor[HTML]{9AFF99} 3,883.11 & \cellcolor[HTML]{9AFF99} 3,883.07 & 0.14 & 3,883.15 & 3,883.18 & 0.11 & 0.05 & \checkmark &  0.83 (Large) \\
Dict comprehension & 13,248.52 & 13,248.54 & 0.10 & 13,248.50 & 13,248.51 & 0.10 & - & - & 0.30 (Small) \\
Generator expression & \cellcolor[HTML]{9AFF99} 19.00 & \cellcolor[HTML]{9AFF99} 19.03 & 0.13 & 3,882.95 & 3,882.97 & 0.11 & 3,863.95 & \checkmark & 32,593.93 (Very large) \\
yield & \cellcolor[HTML]{9AFF99} 19.04 & \cellcolor[HTML]{9AFF99} 19.04 & 0.13 & 3,883.17 & 3,883.18 & 0.11 & 3,864.13 & \checkmark & 32,540.98 (Very large) \\
lambda & 3,882.95 & 3,882.97 & 0.10 & 3,882.95 & 3,882.97 & 0.10 & - & - & 0.01 (Very small) \\
collections.defaultdict & 14,337.73 & 14,337.76 & 0.09 & 14,337.73 & 14,337.75 & 0.11 & - & - & 0.10 (Small) \\
collections.deque & 19.00 & 19.05 & 0.13 & 19.00 & 19.04 & 0.13 & - & - &  0.05 (Very small) \\
zip & \cellcolor[HTML]{9AFF99} 5,815.28 & \cellcolor[HTML]{9AFF99} 5,815.30 & 0.10 & 9,297.17 & 9,297.18 & 0.10 & 3,481.89 & \checkmark & 33,629.04 (Very large) \\
itertools.zip\_longest & \cellcolor[HTML]{9AFF99} 5,815.28 & \cellcolor[HTML]{9AFF99} 5,815.30 & 0.11 & 12,780.13 & 12,780.15 & 0.10 & \cellcolor[HTML]{9AFF99} 6,964.85 & \checkmark &  65,805.65 (Very large) \\ \bottomrule
\end{tabular}
}
\end{table*}

\textit{9. itertools} consist of several functions such as zip\_longest, starmap, tee, and groupby.

\begin{lstlisting}[language=Python,
caption={},
label=code:userhome]
# Note: n is integer
#       a_list is list of n integer elements
#       b_list is list of n/2 integer instances
from itertools import zip_longest
for a, b in zip_longest(a_list, b_list, fillvalue = -1):
    result = min(a, b)
\end{lstlisting}

\section{ Experiment Setup}
We now describe our experiment setup used for the study.
As aforementioned, we provide scripts and datasets as a replication package. 

\subsection{{Tools for measurements}}
For our experiment environment, we implemented a framework to measure memory usages and execution times.
Using the \textit{memory-profiler} library \citep{Web:mem-profiler}, we measured the full memory usage of a Python process with the execution timestamp.

For the environment specification, we used a single machine with AMD Ryzen Threadripper 3970X CPU and 128 GB of DDR4 RAM.
Note that our framework was executed on Ubuntu 20.04 operating system with CPython version 3.9.7 and used only a single process per execution.

\subsection{{Input Parameters}}
For the experiment, we performed at total of 9 runs, each time increasing the input size of the computation.
Hence, for each of the nine Pythonic and non-Pythonic code, we input a list of integer instances from 1 and multiply it by 10 up to 10$^8$ (i.e., in total 9 different input sizes).
Each run was labeled as  \inputSize~= 1 to 9 respectively.
This way, the highest setting is 100,000,000 of random numbers in the single list.
Due to the fact that each run could be non-deterministic and other factors that may impact the run time, we repeat each run experiment for 100 times.

To calculate the improvements, we find the difference in medians of both memory usage and execution time for each run.
Note that we only report if there is improvement over the non-Pythonic code.
Overall, we find that the improvements in both megabytes (MB) and seconds increase at the runs increase, as is shown by Table \ref{tab:mem_time_diff}.

\begin{tcolorbox}
\textbf{Observation 1}:
Results indicate that performance differences become apparent at higher input computations.
\end{tcolorbox}

\subsection{{Data Analysis}}
For the rest of the analysis, we focus on the highest run setting (i.e., Run = 9).
We then report the median and mean memory (in megabytes MB) and execution times (in seconds).

In order to statistically validate the differences between the performance of Pythonic and Non-pythonic codes, we apply an independent sample t-test \cite{Ross:2017}.
We test the null hypothesis that ``\textit{the performance of Pythonic and Non-pythonic codes are the same}''.
The parameters are set the effect size using Cohen's d which indicate differences between two means based on the standard deviation of the population. 
The interpretation is listed as follows: (1) 0 $\leq$ d \textless~0.1 as Very small, (2) 0.1 $\leq$ d \textless~0.35 as Small, (3) 0.35 $\leq$ d \textless~0.65 as Medium, (4) 0.65 $\leq$ d \textless~0.9 as Large, or (5) d $\geq$ 0.9 as Very large.
For a statistical test, we use NumPy \citep{Numpy:2020}, SciPy \citep{SciPy:2020}, and researchpy \citep{Web:researchpy}.

\section{Analysis 1 - Memory Consumption}

\begin{table*}[h]
\centering
\caption{Summary statistics and comparison of execution times (in seconds) between Pythonic and Non-pythonic codes. This result is taken form \inputSize~= 9 (i.e., 10$^8$ integer instances). $\Delta$ is improvement of Pythonic code over Non-Pythonic code (based on median).}
\label{tab:result_time_stat}
\scalebox{0.82}{
\begin{tabular}{@{}crrrrrrccc@{}}
\toprule
 & \multicolumn{3}{c}{\textbf{Pythonic}} & \multicolumn{3}{c}{\textbf{Non-Pythonic}} & \multirow{2}{*}{\textbf{$\Delta$}} & \multirow{2}{*}{\textbf{t-test}} & \multirow{2}{*}{\textbf{Cohen's d}} \\ \cmidrule(r){2-7}
 & \multicolumn{1}{c}{\textbf{Median}} & \multicolumn{1}{c}{\textbf{Mean}} & \multicolumn{1}{c}{\textbf{SD}} & \multicolumn{1}{c}{\textbf{Median}} & \multicolumn{1}{c}{\textbf{Mean}} & \multicolumn{1}{c}{\textbf{SD}} & & (p-value $<$ 0.001)  &  \\ \midrule
List comprehension & \cellcolor[HTML]{9AFF99} 3.04 & \cellcolor[HTML]{9AFF99} 3.04 & 0.02 & 8.82 & 8.85 & 0.24 & 5.78 & \checkmark & 34.14 (Very large) \\
Dict comprehension & \cellcolor[HTML]{9AFF99} 10.98 & \cellcolor[HTML]{9AFF99}  11.01 & 0.16 & 15.44 & 15.45 & 0.32 & 4.46 & \checkmark & 17.24 (Very large) \\
Generator expression & \cellcolor[HTML]{9AFF99} 6.19 & \cellcolor[HTML]{9AFF99} 6.23 & 0.17 & 7.29 & 7.3 & 0.09 & 1.10 & \checkmark & 7.58 (Very large) \\
yield & \cellcolor[HTML]{9AFF99} 7.06 & \cellcolor[HTML]{9AFF99} 7.12 & 0.22 & 8.76 & 8.76 & 0.06 & 1.70 & \checkmark & 10.28 (Very large) \\
lambda & \cellcolor[HTML]{9AFF99} 11.99 & \cellcolor[HTML]{9AFF99} 11.99 & 0.17 & 12.52 & 12.51 & 0.14 & 0.53 & \checkmark & 3.36 (Very large) \\
collections.defaultdict & 25.41 & 25.53 & 0.56 & \cellcolor{yellow} 18.57 & \cellcolor{yellow} 18.65 & 0.38 & - & \checkmark & 14.28 (Very large) \\
collections.deque & \cellcolor[HTML]{9AFF99} 12.33 & \cellcolor[HTML]{9AFF99} 12.32 & 0.09 & 14.4 & 14.41 & 0.11 & 2.07 & \checkmark & 20.95 (Very large) \\
zip & \cellcolor[HTML]{9AFF99} 15.8 & \cellcolor[HTML]{9AFF99} 15.79 & 0.37 & 26.82 & 26.85 & 0.45 & 11.02 & \checkmark & 26.85 (Very large) \\
itertools.zip\_longest & \cellcolor[HTML]{9AFF99} 28.18 & \cellcolor[HTML]{9AFF99} 28.39 & 0.89 & 60.44 & 60.44 & 0.91 & \cellcolor[HTML]{9AFF99} 32.25 & \checkmark & 35.36 (Very large) \\ \bottomrule
\end{tabular}
}
\end{table*}

Table \ref{tab:result_mem_stat} shows the summary statistics of  memory usage for each Pythonic and Non-pythonic code at the highest input points (size is in \inputSize~= 9).
Under these settings, we find that five out of nine Pythonic idioms had significant memory performance improvements when compared to the Non-pythonic counterparts (i.e., highlighted in green, statistically significant difference and have large effect size).
The Pythonic idiom that peak performance is List comprehension, Generator expression, yield, zip and itertools.zip\_longest.
We also find that using itertoools.zip\_longest can save more memory than other idioms (i.e., a median difference of 6,965.85 MB, an effect size is 65,805.65 standard deviations or very large).

On the other hand, there is no significant difference from Dict comprehension, lambda (equally 3,882.95 MB), collections.defaultdict (equally 14,337.73 MB) and collections.deque (equally 19.00 MB).
We believe that this is because both idioms use a multiple iterators in parallel, so it does not allocate extra space for pairing those items in the lists.
\begin{tcolorbox}
\textbf{Observation 2}:
Five out of nine Pythonic idioms statistically significantly consume less memory than their Non-pythonic counterparts.
(i.e., 
List comprehension, dict comprehension, generator expression, yield, lambda, zip and itertools.zip\_longest
)
\end{tcolorbox}

\section{Analysis 2 - Execution Time}
Table \ref{tab:result_time_stat} shows summary  statistics for execution times at the highest setting (\inputSize~= 9).
Here we find that eight out of nine Pythonic idioms improve the execution times (i.e., statistically significant difference and have very large effect size are highlighted in green in the Table).
The Pythonic idioms are List comprehension, Dict comprehension, Generator expression, yield, lambda, collections.deque, zip and itertools.zip\_longest.
We can see that using Pythonic idiom can save up to 32.25 seconds (i.e., from itertools.zip\_longest).

Also from Table \ref{tab:result_time_stat}, we find that  collections.defaultdict took more execution times than Non-pythonic code style (i.e., 25.41 seconds \textgreater~18.57 seconds, highlighted in yellow).
As stated in the official Python document, defaultdict is a subclass of the built-in dict class.
We suspect that the new mechanism of defaultdict causes the slower speed than the built-in dict, even though it helps developers write a shorter code.
Note that with our experiment setup, the time differences are still within one minute.

\begin{tcolorbox}
\textbf{Observation 3}:
Eight out of nine idioms were faster in execution than the Non-pythonic counterparts at scale.
The only exception was the collections.defaultdict idiom.
\end{tcolorbox}


\section{Implications and Future Challenges}
We now discuss the implications and challenges for our research directions.

\subsection{Limitations}
A key limitation of this work is that our code examples may not reflect what is practiced in real-world application.
Furthermore, we are unsure the extent to which we could find and replace the non Pythonic code with their Pythonic counterparts. 
Another key limitation is the execution environment. 
Currently our setup is using a high-powered machine. 
This might not be realistic in a setting where the user may use a lower specification machine or have less available memory.
We also need to consider whether or not these Pythonic codes are prevalent in the wild.
Finally, in this study, we only focus on nine Pythonic examples.
There may be other forms of Pythonic writing that are not covered by this research.
We envision that all these limitations open up future avenues for research.

\subsection{Implications}
The main takeaway message for the paper is that writing Pythonic code does matter at scale.
We are not sure to what extent that this may relate to real-world examples but shows that it is worth exploring in the future.

\textit{For Python programmers,} our results are encouraging for data scientist and other Python programmers that deal with large computations and limited memory and time. 
Furthermore, as a community, it is another reason to encourage newcomers to exploit the benefits of writing the Pythonic way.
For instance, it is worthwhile to incorporate list comprehension, generator expression, zip and itertools.zip\_longest as they save both time and memory.
One example of a set of Python programmers would be the maintainers of PyPI libraries that deal with large datasets.

\textit{For researchers,} based on these results similar to \cite{Phan-udom:ICSME2020}, tool support could be used to assist with suggestions to change non-Pythonic to their Pythonic versions.
The tool could be helpful for novices, or for developers that programming in multiple languages and are unaware of the Pythonic way.

\textit{For educators,} it would be useful to introduce such idioms to students that would like to fully maximize their usage of the language.
We suggest that this could be part of the more advanced level of usage of the Python programming language.

\subsection{Challenges and Future Directions}
We identify three key challenges for future research directions.
The first challenge is to detect and retrieve Pythonic idioms in the wild.
The second challenge for analyzing Python code in the wild is automatic compiling real-world code.
The final challenge is to evaluate whether or not the savings in performance makes a difference in that certain scenario.
In this case, we would have to conduct a user study to get developer perspectives.
Also, we would have to focus on specific tasks such as machine learning models to concretely evaluate the benefits.
Apart from analyzing source code of software projects, we also would like to explore other sources such as the Kaggle dataset for data scientists.

\section*{Acknowledgment}
This work has been supported by Japan Society for the Promotion of Science KAKENHI Grants Grant Number JP20H05706 and JP20K19774.

\bibliographystyle{ACM-Reference-Format}
\bibliography{bibliography}

\end{sloppy}
\end{document}